\documentclass[superscriptaddress,secnumarabic,amssymb,amsmath,nobibnotes,aps,prd,showkeys,showpacs,nofootinbib,preprint]{revtex4-1}

\usepackage{graphicx}
\usepackage{float}
\usepackage{bm}
\usepackage{amsmath}
\usepackage{amsfonts}
\usepackage{amssymb}
\usepackage{epstopdf}
\usepackage{natbib}%
\newcommand{\bee}{\begin{equation}}
\newcommand{\eee}{\end{equation}}
\newcommand{\eaa}{\end{eqnarray}}
\newcommand{\baa}{\begin{eqnarray}}
\def\ni{\noindent}

\usepackage{color}

\begin{document}

\title{\Large Barrow's black hole entropy and \\ the equipartition theorem}

\author{Everton M. C. Abreu}\email{evertonabreu@ufrrj.br}
\affiliation{Departamento de F\'{i}sica, Universidade Federal Rural do Rio de Janeiro, 23890-971, Serop\'edica, RJ, Brazil}
\affiliation{Departamento de F\'{i}sica, Universidade Federal de Juiz de Fora, 36036-330, Juiz de Fora, MG, Brazil}
\affiliation{Programa de P\'os-Gradua\c{c}\~ao Interdisciplinar em F\'isica Aplicada, Instituto de F\'{i}sica, Universidade Federal do Rio de Janeiro, 21941-972, Rio de Janeiro, RJ, Brazil}

\author{Jorge Ananias Neto}\email{jorge@fisica.ufjf.br}
\affiliation{Departamento de F\'{i}sica, Universidade Federal de Juiz de Fora, 36036-330, Juiz de Fora, MG, Brazil}

\author{Ed\'esio M. Barboza Jr.}\email{edesiobarboza@uern.br}
\affiliation{Departamento de F\'isica, Universidade do Estado do Rio Grande do Norte, 59610-210, Mossor\'o, RN, Brazil}




\begin{abstract}
\ni  Barrow's entropy appears from the fact that the black hole surface can be modified due to the quantum gravitational 
outcome.   The measure of this perturbation is given by a new exponent $\Delta$.  
In this letter we have shown that, from the standard mathematical form of the equipartition theorem, we can relate it to Barrow's entropy.  
After that, we tested the thermodynamical coherence of the system by calculating exactly the heat capacity which established an interval of the possible thermodynamical coherent values of Barrow entropic exponent.
\end{abstract}
\pacs{04.70.Dy, 98.80.-k}
\keywords{Barrow entropy, equipartition theorem}

\maketitle

Black holes (BH) are one of the most intriguing and fascinating astronomical entities that encompasses the main issues of quantum gravity analysis.  The origin of a BH, i.e., a collapsing star, for example, makes us to realize that the geometry of short distance fluctuations can be improved until macroscopic distances.   This connection between high-energy and low-energy physics can involve serious discussions about the dynamics of these systems at Planck scale.   And it has consequences concerning low-energy experiments.

Having said that, it is well known at this juncture, that there exists an association between BH physics and thermostatistical formalisms. 
 As an example, we can compute the specific temperature and entropy of a BH.   Both rely on the BH horizon  \cite{gh}. Consequently, it was suggested an extension of this computation, the so-called {\it thermodynamics of spacetime}, where we can use the concepts of thermodynamics at the horizon of the Universe, see \cite{jpp} and references therein for details.

Besides, working on the thermodynamics first law at the apparent horizon, we can obtain the Friedmann equations.  It can also be shown that from the Friedmann equations  we can obtain the first law \cite{varios-1}. These precise calculations show that they are efficient both in general relativity (GR) as well as  in several  modified gravity approaches, although in the latter formalisms the entropy relation is in general altered \cite{varios-2}.  

We can also think of  the  second law of thermodynamics, which, considering BH physics, it  was  extended to the concept of {\it generalized second law of  thermodynamics}, i.e., where the standard entropy combined with the BH horizon entropy is an increasing, or stable at least (non-decreasing), time function \cite{bu}. We can also use this concept when consider the Universe horizon, where  the total entropy of the interior of the Universe together with the entropy of its horizon can be  a non-decreasing time function \cite{barrow}. We can demonstrate that this idea is always true for GR, although it is not always what happens in  
modified theories of gravity. Therefore, we can employ it to obtain its constraints \cite{varios-3}.

A few days back, Barrow \cite{barrow-2}   analyzed the circumstance where quantum gravitational effects could cause about intricate, fractal structure on 
the BH surface.  It changes its actual horizon area, which in turn leads us to a new BH entropy relation, namely,
\bee
\label{barrow}
S_B\,=\,\bigg(\frac{A}{A_o} \bigg)^{\frac \Delta2 +1}, 
\eee

\ni where $A$ is the usual horizon area  and $A_o$ the Planck area.   It is important to say that  this extended entropy is different from the 
standard ``quantum-corrected" one \cite{saridakis-1} with logarithmic corrections \cite{kc}, although it is a kind of Tsallis nonextensive entropy expression \cite{varios-4}. We can see clearly that this quantum gravitational perturbation is represented by the new exponent $\Delta$.  There are some characteristic values for $\Delta$.  For example, when $\Delta=0$  we have the simplest horizon construction.  In this case we obtain the 
well known  Bekenstein-Hawking entropy.   On the other hand, when   $\Delta=1$ we have the so-called  maximal deformation.   

In this letter we are interested specifically in the analysis of Barrow's exponent by using the compatibility of Barrow's entropy with the equipartition law.  We have also tested the validity of our result with the thermodynamical coherence of the model by calculating its heat capacity.
To accomplish the task, we will use thermodynamic functions, which are normally used in BHs physics, they are entropy and temperature. Throughout this letter we will use $\hbar=c=k_B=1$. In the context of the usual BH area entropy law, $S=A/4G$.  We will also assume that the number $N$ of degrees of freedom (dof) of the horizon satisfy the standard equipartition law \cite{pad}
\baa
\label{equi}
M= \frac{1}{2} N T \,,
\eaa

\ni where $T$ is the temperature and $M$ is the BH mass. 



The thermodynamics of BHs were constructed upon the basis of the ideas of both the entropy
and temperature of BH \cite{jdb,swh,wald}. The temperature of a BH horizon is directly proportional to
its surface gravity. In Einstein gravity theory, the BH horizon entropy is proportional to its
horizon area, i.e., the BH entropy area law.

Our main target will be the Schwarzchild BH entropy, which will depict the horizon.   Following Barrow deformed entropy, given in Eq. \eqref{barrow} for BHs \cite{barrow} it is given by
\bee
\label{a1}
S_B\,=\,\Bigg(\frac{A}{4G}\Bigg)^{\frac \Delta2 + 1}\,\,,
\eee

\ni where $G$ is the gravitation constant, $4G$ is the Planck area and $A$ is the standard horizon area.   The new exponent, $\Delta$, is responsible for the quantum gravitational deformation \cite{saridakis-2}.   There are two characteristic values for $\Delta$.   Namely, $\Delta=0$ shows the well known Bekenstein-Hawking entropy, which is the easiest horizon framework.   For $\Delta=1$, we have a fractal framework.  So, the entropy in Eq. \eqref{a1} connects the BH area entropy to the area $A$ of the horizon.  In BH physics, the area $A$ of the horizon can be associated with the source mass $M$ through the relation
\bee
\label{a2}
A=16 \pi G^2 M^2 \, .
\eee

\ni It will be assumed that the number of degrees of freedom (dof), $N$, of the horizon obey the standard equipartition theorem \cite{pad}
\bee
\label{a3}
M= \frac{1}{2} N T \,,
\eee

\ni where $T$ is the temperature and $M$, the mass of the BH.   Substituting the area in Eq. \eqref{a2} into Eq. \eqref{a1}, we have that
\bee
\label{a4}
S_B\,=\,\Bigg(\frac{16\pi G^2 M^2}{4G}\Bigg)^{\frac \Delta2+1} \qquad \Longrightarrow \qquad S_B \,=\,\Big(4\pi G M^2 \Big)^{\frac \Delta2+1}\,\,.
\eee

The temperature is given by
\bee
\label{a5}
\frac{1}{T}=\frac{\partial S(M)}{\partial M} \,,
\eee

\ni and using Eq. \eqref{a4}
\bee
\label{a6}
\frac 1T \,=\,  \Big( \Delta + 2 \Big) \Big(4\pi G\Big)^{\frac \Delta2+1} M^{\Delta +1}\,\,.
\eee

\ni We will use that the number of dof, $N$, in the horizon can be obtained by \cite{ko} 
\bee
\label{a7}
N\,=\,4\,S\,\,,
\eee

\ni where $S$ is the specific entropy that describes the horizon.   Hence, using Eqs. \eqref{a4} and \eqref{a7} we have that
\bee
\label{a8}
\frac N4 \,=\, \Big(4\pi G\Big)^{\frac \Delta2+1} M^{\Delta+2}\,\,,
\eee

\ni but, substituting the result in Eq. \eqref{a6} for the temperature into Eq. \eqref{a8}, we have that
\bee
\label{a9}
M\,=\,\frac 12 \bigg( 1\,+\,\frac \Delta2 \bigg) N\,T\,\,,
\eee

\ni which corresponds to the horizon energy in Barrow's entropy model. From the last equation we can notice the appearance of an extra term $\Delta/2$ in the usual equipartition theorem, Eq. \eqref{a3}. When we make $\Delta = 0$ we recover the usual equipartition law.



Let us discuss now the physical coherence of this structure.   We can do that by calculating the heat capacity of the model.   In other words, the sign of the heat capacity can pinpoint the stability condition of BHs.   Namely, the heat capacity must be positive for stable thermodynamical systems.   A negative heat capacity indicates that there is a violation of the laws of thermodynamics, i.e., thermodynamical unstableness.

The heat capacity is given by
\bee
\label{a13}
C\,=\,-\,\frac{[S_{BH}^{\prime}(M)]^2}{S_{BH}^{''}(M)} \,\,,
\eee

\ni where the prime means a derivative relative to $M$.   So, substituting Barrow's entropy in Eq. \eqref{a4} into Eq. \eqref{a13} 
we have that
\bee
\label{a15}
C_B\,=\,-\,\frac{\Big[\,\Big(4\pi G\Big)^{\frac \Delta2+1}\Big(\Delta + 2\Big)\Big]}{\Delta\,+\,1}\,M^{\Delta+2} \,\,,
\eee

\ni which means that, for stability, we must have that
\bee
\label{a16}
\frac{\Delta+2}{\Delta + 1} < 0 \qquad \Longrightarrow \quad -2<\Delta<-1 \qquad \Longrightarrow \quad C_B > 0 \,\,.
\eee

\ni Hence, for thermodynamical coherence with the equipartition law, we have that $\Delta$ has to be in the interval $-2<\Delta<-1$.   
But the interval for $\Delta$ is $0 \leq \Delta \leq 1$ \cite{barrow-2}.   Therefore, Barrow's BH is unstable, as it is expected.

For $\Delta=0$ in Eq. \eqref{a15}, for a smooth spacetime structure, we have that 
\bee
\label{h-capacity-BH}
C_{B}\,=\,-\,8 \pi G M^2 \,\,,
\eee

\ni which reproduces the usual value of the heat capacity of a BH and it means, as well known, that a BH is thermally unstable.    The negative heat capacity in this regime means that a slight drop in BH's temperature will cause an extra drop as the energy keeps being absorbed. The process will continue forever and the BH will keep feeding on the surrounding heat bath. Moreover, a slight rise in BH's temperature from the equilibrium value will cause the BH to radiate some net energy.   Hence, it will increase its temperature still further.   It will, eventually, conduct to the explosive vanishing of the BH altogether.

For $\Delta=1$ in Eq. \eqref{a15} we have that 
\bee
\label{h-capacity-BH}
C_{B}\,=\,-\,12 \Big(\pi G\Big)^{3/2} M^3 \,\,.
\eee

\ni which corresponds to the heat capacity of a maximal deformation of spacetime, i.e., for the most intricate. 


To conclude, we can mention that the Barrow's entropy originates from the fact that the BH surface can be perturbed by the so-called quantum gravitational effects.  We can measure its deviation from the Bekenstein-Hawking entropy through a new exponent $\Delta$, where $\Delta = 0$ means Bekenstein-Hawking entropy, and $\Delta=1$ means the most intricate case.  In this work we  have calculated precisely the expression of the equipartition law which corresponds to the horizon energy in Barrow's entropy model.  After that,  to observe the application of the thermodynamical coherence of the model, we have calculated the heat capacity of the system, which must be a positive quantity.  It brought a negative interval of validity of Barrow's entropy exponent.   However, since the interval known for the validity of $\Delta$ is positive, it means that the Barrow's black hole is unstable, as it is expected.


\section*{Acknowledgments}

\ni The authors thank CNPq (Conselho Nacional de Desenvolvimento Cient\' ifico e Tecnol\'ogico), Brazilian scientific support federal agency, for partial financial support, Grants numbers  406894/2018-3 (E.M.C.A.) and 303140/2017-8 (J.A.N.).

\end{document}